\documentclass[preprint,authoryear,12pt]{elsarticle}

\usepackage{amssymb}

\journal{Astroparticle Physics}

\begin{document}

\begin{frontmatter}



\title{On the Origin of the 6.4 keV Line in the Galactic
Center Region}


 \author[label1,label2]{V. A. Dogiel\footnote{Tel.: +7 499 132 6235; Fax: +7 499 135 8533; E-mail address: dogiel@lpi.ru} }
\address[label1]{I.E.Tamm Theoretical Physics Division
of P.N.Lebedev Institute of Physics, Leninskii pr. 53, 119991
Moscow, Russia}
\address[label2]{Moscow Institute of Physics and Technology, 141700 Moscow Region, Dolgoprudnii, Russia}
 \author[label1]{D. O. Chernyshov}
 \author[label1]{A. M. Kiselev}
 \author[label3]{and K.-S.Cheng}
\address[label3]{Department of Physics, University of
Hong Kong, Pokfulam Road, Hong Kong, China}

\begin{abstract}
We analyse  the 6.4 keV iron line component produced  in the Galactic Center (GC) region by cosmic rays in dense molecular clouds (MCs) and in the diffuse molecular gas. We showed that this component, in principle, can be seen in several years in the direction of the cloud Srg B2. If this emission is produced by low energy CRs which ionize the interstellar molecular gas the intensity of the line is quite small, $<1$\%. However, we cannot exclude that local sources of CRs or X-ray photons nearby the cloud may provide much higher intensity of the line from there. Production of the line emission from molecular clouds depends strongly on processes of CR penetration into them. We show that turbulent motions of neutral gas may generate strong magnetic fluctuations in the clouds which prevent free penetration of CRs into the clouds from outside.
We  provide a special analysis of the line production by high energy electrons. We concluded that these electrons hardly provide the diffuse 6.4 keV line emission from the GC because their density is depleted by ionization losses. We do not exclude that local sources of electrons may provide an excesses
of the 6.4 keV line emission in some molecular clouds and even
reproduce a relatively short time variations of the iron line emission. However, we doubt whether a single electron source provides the simultaneous short time variability of the iron line emission
from clouds which are distant from each other on hundred pc as observed for the GC clouds. An alternative speculation is that local electron
sources could also provide the necessary effect of the line variations in
different clouds that are seen simultaneously  by chance  that seems, however, very unlikely.
\end{abstract}

\begin{keyword}
Galaxy: center --- ISM: clouds --- cosmic rays --- line: formation
--- X-rays: ISM


\end{keyword}

\end{frontmatter}


\section{Introduction}
The origin of 6.4 keV emission from molecular clouds (MCs) of the
Galactic Center (GC) has been discussed from 1993
\citep[see][]{suny}, and it was
 concluded that this emission was produced by keV photons
emitted by Srg A* about 100 yr ago \citep[see e.g. last
publications of][and references therein]{inu09,ponti10,terrier,
nob11,capelli12,ryu12,clavel13}. This line emission was observed
first in the direction of the cloud Srg B2 \citep{koya1}. The
subsequent observations of the line and X-ray continuum emission
from this cloud found its prominent time variability whose time
characteristic corresponded to the period for which the front of
primary photons from Sgr A* crossed the cloud \citep[see
e.g.][]{inu09,terrier}. Analysis of the Chandra data provided by
\citet{clavel13} showed that this activity of SgrA* can be
presented as short X-ray flares whose duration is about 10 yr. In
principle, past flares of Sgr A* can be found on time scales of
ten thousand years \citep[see][]{sun02}

For the last ten years the 6.4 keV intensity of Sgr B2 has dropped
down in about three times in comparison with its maximum value in
2000 \citep[see][]{nob11}.  The background 6.4 keV emission for
e.g. the cloud Sgr B2 is expected to be seen  in several years
when the front of primary Sgr A* photons leaves finally the cloud.
Does it mean that in several years the line flux from Sgr B2 will
drop down to zero?

The  6.4 keV emission from the GC clouds can also be provided  by
cosmic rays (CRs). Three aspects of the alternative iron line
production by CRs have to be clarified:
\begin{itemize}

\item It is known that the GC medium is filled by relatively low energy CRs. As
\citet{indrio10} showed,  CRs of MeV energies  ionize the
interstellar gas  and are absorbed there. The density of these CRs
in the GC can be derived from measurements of the absorption line
of the ionised molecular hydrogen ($H_3^+$). Thus, \citet{oka05}
and \citet{goto08,goto11} found an unusually high ionization rate
in the GC, which is not observed in other parts of the Galactic
Disk. The ionization rate, $\zeta$,  is almost uniform throughout
of the GC on scales about 200 pc, $\zeta \sim (1-3)\times
10^{-15}$ s$^{-1}$. This suggests  a single and widespread
mechanism of ionization there. Then the question is what is the
 level of the 6.4 keV line emission produced by these
MeV CRs. Alternatively, the background flux of the 6.4 keV line in
the GC can also be provided by a hypothetic injection of
subrelativistic protons  by a star accretion onto the central
black hole \citep[see][]{dog09}. They estimated this flux for the
molecular cloud Sgr B2 and found that in several years it might be
about 15\% of the maximum observed in 2000. Of course, these
predictions are strongly model dependent and cannot be considered
as reliable;
\item The line emission from several clouds like the
Arches cluster region, Sgr C, G0.162-0.217, GO.11-0.11  and others
\citep[see][]{fuk09,tsuru,tati12,yus3} is  unlikely due to
photoionization  because e.g. some of them  do not show any time
variability as expected in the photoionization model. The question
is whether the line flux from these clouds is produced by
background CRs  or by CRs from local sources;
\item Although the photon origin of the 6.4 keV line
emission is widely accepted, \citet{yus3} developed a model of the
line production by relativistic electrons which explained also
 the flux time variability  from the GC clouds.
\end{itemize}

Below we  provide investigations of the iron line generation by CRs in
the GC.

\section{6.4 keV Emission from Sgr B2 Produced by Subrelativistic Protons }
We investigate first the case of  ionization by protons. As we
noticed above CRs ionize the molecular gas in the GC with the rate
$\zeta \geq 3\times 10^{-15}$ s$^{-1}$ which is almost constant in
the  GC region. The rate of ionization by subrelativistic protons
reads as
\begin{equation}
 \zeta \simeq \int\limits_{I({\rm H_2})}^{E_p^{\rm max}}dE_p
\sigma_{p}^{\rm ioni}(E_p)  n_{\rm p}(E_p)v_p(E_p)    ~, \label{ion_p}
\end{equation}
where $\sigma_{p}^{\rm ioni}$ is the cross-sections of ionization
by protons  taken e.g. from \citet{tati03}, $n_{\rm
p}(E_p)$ is the spectrum of primary protons, $I({\rm H_2})$ is the ionization potentials of hydrogen, $E^{\rm max}_p$ is the maximum energy of
primary protons,   and $v_p$ is the velocities of primary
protons. As in \citet{dog13} the contribution from knock-on electrons is neglected.

\begin{figure}
\centering
\includegraphics[width=0.6\textwidth]{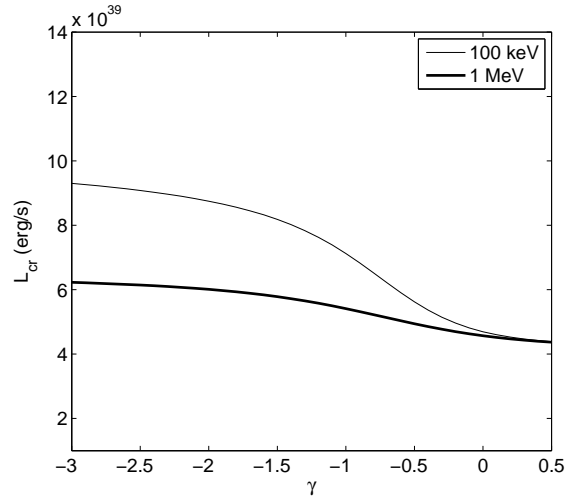}
\caption{ The luminosity of CR protons  in the GC region,$L_{cr}$,
necessary to provide the ionization rate $\zeta = 3\times
10^{-15}$ s$^{-1}$ as a function of the proton spectral index
$\gamma$. Thick line: for the range of protons 1 - 100 MeV,
thin line: the range of protons 100 keV - 100
MeV}\label{p_lum}
\end{figure}

Parameters of the proton spectrum necessary for the observed
ionization rate in the GC, $\zeta = 3\times 10^{-15}$ s$^{-1}$,
can be derived, if we take the proton spectrum as power-law,
$n_p(E_p)=KE_p^\gamma$, where $E_p$ is the kinetic energy of protons, $E_p=\sqrt{p^2c^2+(m_pc^2)^2}-m_pc^2$. In Fig. \ref{p_lum} we presented the
luminosity of protons in the GC region of the radius 200 pc and
thickness 60 pc derived from Eq. (\ref{ion_p}).
We see that the required proton luminosity in the GC region, $\leq
10^{40}$~erg~s$^{-1}$,  seems to be quite reasonable if supplied
by SNs in the GC \citep[see][]{crock11}. As
\citet{cheng07,cheng11} and \citet{dog09b} showed, this luminosity
of subrelativistic and relativistic CRs in the GC  can also be
provided by processes of star accretion onto the central black
hole, though as we mentioned these estimates cannot be considered
as reliable.

\begin{figure}
\centering
\includegraphics[width=0.67\textwidth]{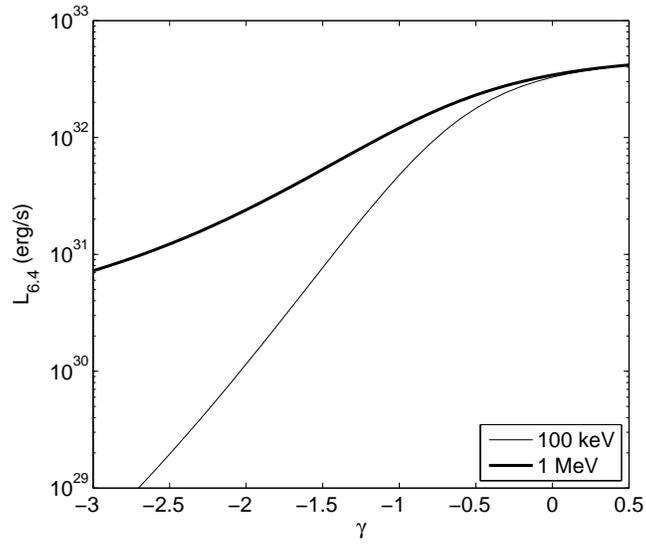}
\caption{6.4 keV flux from Sgr B2 generated by protons as a
function of their spectral index $\gamma$. The spectrum parameters
are the same as in Fig. \ref{p_lum}.}\label{Sgr_p}
\end{figure}

Then for the proton spectrum derived from Eq. (\ref{ion_p}) we can estimate an expected background level of the 6.4 keV
flux from Sgr B2 provided by the protons when the front of Sgr A*
photons leaves the cloud. The equation is
\begin{equation}
F_{6.4} \simeq \eta_{Fe}n_HV{\int\limits_{I({\rm Fe})}^{E_p^{\rm
max}}}dE_p \sigma_{p}^{\rm Fe}(E_p)  n_{\rm p}(E_p)v_p(E_p)
\end{equation}
where $\sigma_{p}^{\rm Fe}$ is the cross-section of iron
ionization  by protons \citep[see][]{tati03}, $\eta_{Fe}$ is the
iron abundance,  $I({\rm Fe~K})=7.1$ keV is the ionization
potentials of iron, and $V$ is the cloud volume. The total mass of
Sgr B2 is poorly known.  In the volume  of  42 pc diameter this
mass ranges from $2\cdot 10^5$ to $7\cdot 10^6$ M$_\odot$
\citep{oka}. For estimates we take the mass of $10^6$ M$_\odot$
and the fixed ionization rate in the GC $\zeta = 3\times 10^{-15}$
s$^{-1}$. The background flux of the 6.4 keV line as a function of
proton spectral index $\gamma$ and the minimum energy of protons,
$E_{min}$, equaled 1 MeV (thick solid line) and 100 keV (thin
solid line) is shown in Fig. \ref{Sgr_p}.
\begin{figure}
\centering
\includegraphics[width=0.67\textwidth]{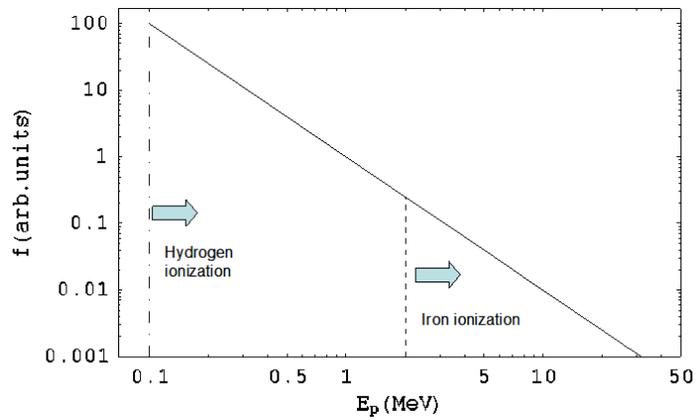}
\caption{The energy ranges of subrelativistic protons which ionize
the hydrogen gas and iron atoms.}\label{fig}
\end{figure}
 For the fixed ionization rate $\zeta$ the expected 6.4 keV flux depends on the spectral index of protons and the cut-off energy of protons $E_{min}$, the steeper is the spectrum and the smaller is $E_{min}$, the less is the expected flux of the diffuse 6.4 keV flux from the GC region. This effect is illustrated in  Fig. \ref{fig} where we showed the energy range of subrelativistic
protons which ionize the hydrogen gas and iron atoms \citep[see
for the ionization cross-sections][] {tati03,dog13}.

   From Fig. \ref{Sgr_p} we can conclude that for steep spectra of
protons we expect further decrease of the flux which will reach
the level below $\leq 1$\% of its maximal value measured in 2000,
if there are no local sources of CRs in the Sgr B vicinity. For
$\gamma<0.5$ the flux of background 6.4 keV line emission from Sgr
B2 depends strongly on $E_{min}$ of the protons. We see from the
figure that this flux is negligible if $E_{min}=100$ keV. This
conclusion is in a full agreement with results obtained by
\citet{dog13} (see Fig. 1 in that paper).

A relatively high flux of the 6.4 keV emission from Sgr B2  may
mean that accretion processes generate, indeed,  high energy
subrelativistic protons in the GC as assumed by \citet{dog09}. In
this case  the spectrum of protons is extremely hard, $\propto
E^{0.5}$, \citep[see][]{dog09a} providing a more effective
ionization of iron  than in the cases of steep proton spectra (see
Fig. \ref{Sgr_p}). Alternatively, there may be
 local sources of CRs or X-ray photons nearby Sgr B2.

\section{Structure of magnetic fields inside the clouds and processes of CR propagation there}
Although the model of photoionization by Sgr A* photons is widely
accepted several molecular clouds in the GC do not fit with this
interpretation, e.g. Arches cluster region, G0.162-0.21, GO.13-013 and
others. Their fluxes are not time variable and the equivalent
width of the 6.4 kev line does not correspond to ionization by
photons \citep[see, e.g.][]{fuk09,ponti10,yus3}. A nice example is
the molecular cloud nearby the Arches cluster  whose 6.4 keV
emission may be generated by CRs. Its emission was analysed in
details by \citet{tati12}, who showed that it was most likely
produced by subrelativistic ions. Other cases were investigated by
\citet{goto13}  who assumed that molecular clouds in the central
few parsecs region of the Galaxy are ionised by protons emitted by
Sgr A*. \citet{yus4} suggested that  the 6.4 keV flux from the cloud  GO.13-013 is generated
by electrons from local sources.

Estimates of the line emission from a molecular cloud depend
strongly on the processes of CR penetration into the clouds. There are
two limit cases:
\begin{itemize}
\item a) when CRs freely penetrate into the clouds. As
\citet{kuls69} showed the MHD-waves are damped
due to ion-neutral friction. Therefore, e.g. \citet{morf82,yus3} assumed that there
are no MHD-waves for scattering  inside the clouds. Just in this
approximation we obtained above the estimations for Sgr B2;
\item
Another limit case is when a strong magnetic turbulence inside the
clouds excited by chaotic motions  of the neutral gas prevents CR
penetration into the clouds \citep[see][]{dog87,dog05,dog90}. In
this case the spatial diffusion coefficient inside the clouds,
$D_c$, is much smaller than in the intercloud medium  that leads to  a depletion of CR density inside the clouds due to ionization losses and $p-p$ collisions. This effect may be seen from $\gamma$-ray observations of nearby molecular clouds \citep[see][]{neronov} . Then  the CR
distribution in the clouds requires special calculations, as
was done e.g. in \citet{dog09}. Below we discuss this  case in
more detail and take parameters of the Arches complex as an
example.
\end{itemize}

The diffusion coefficient inside the clouds, $D_c$ is determined
by the spectrum of magnetic fluctuations excited there. From
observations it follows that the neutral gas of the clouds is
strongly turbulent. The speed of turbulent motions reaches about
10 km s$^{-1}$. The turbulence has a power-law Kolmogorov-like
spectrum in a very broad range of scales from supersonic
\citet{lars} to subsonic \citet{my} regions \citep[see
also the review of][]{falgarone}:
\begin{equation}
v(L)=1.1~L^\alpha(pc)~\mbox{km s$^{-1}$ where $\alpha\simeq
0.3\div 0.5$} \label{ul}
\end{equation}
where $v$ is the velocity of turbulent motions and $L$ is its
scale ($0.01<L<300$ pc).

Though the ionization degree in the clouds is small, $x_e\leq
10^{-4}$, motions of the neutral gas component generate a
turbulence of the ionized gas by friction, and thus excite
fluctuations of magnetic fields. The system of equations for
fluctuations of the ionized component velocity $u$ and magnetic
fields $B$ is
\begin{eqnarray}
&&\left({{\partial \textbf{u}}\over{\partial
 t}}+(\textbf{u}\cdot\nabla)\textbf{u}\right)={1\over\rho_{\rm i}}\left[-\nabla P_{\rm i}+
 {{\left(\nabla\times
 \textbf{B}\right)\times\textbf{B}}\over{4\pi}}\right]+\nu_{\rm i}\nabla^2\textbf{u}+\nonumber\\
 &&+{{\nu_{\rm i}}\over
 3}\nabla\nabla\cdot\textbf{u}-\mu_{\rm in}(\textbf{u}-\textbf{v})\,,\nonumber\\
&&{{\partial \textbf{B}}\over{\partial
 t}}=\nabla\times(\textbf{u}\times\textbf{B}),~~{{\partial{\rho}}\over{\partial
 t}}+\nabla\cdot(\rho\textbf{u})=0,~~\nabla\cdot\textbf{B}=0\,,
 \label{ihd}
 \end{eqnarray}
 where $v$ is the turbulent velocity of the neutral gas, $\rho_{\rm i}$, $\nu_{\rm i}$ etc. are characteristics of the ionized fraction of
 the gas and $\mu_{\rm in}$ is the frequency of collision between ionized ions and
 neutral hydrogen.

 This set of equations was analysed by \citet{dog87,dog05} who
 showed that the energy of  magnetic field  fluctuations is
 concentrated at small scales ($L_{\nu}\sim 10^{13}$ cm) where dissipation
 processes are essential. In this case propagation of magnetized
 relativistic charged particles along spaghetti-like magnetic field lines can be
 described as diffusion with the coefficient $D_c\sim \pi cL_\nu/6$.

Later  \citet{kiselev13} provided a more accurate analysis of
these equations and took into account the influence of a large
scale magnetic field onto the spectrum of magnetic turbulence.
They assumed that magnetic field consists of a large scale average
field and small scale magnetic fluctuations ${\bf B}= \textbf{B}_0
+ \textbf{b} $ and $ \textbf{B}_0 = {const}$. For the correlator
of magnetic fluctuations
\begin{equation}
\langle b_i(\textbf{x},t)b_j(\textbf{x+r},t) \rangle = 2 Q(r)
\delta_{ij} + r Q'(r) \left(\delta_{ij} - \frac{r_i r_j}{r^2} \right)
\end{equation}
 \citet{kiselev13} derived the equation which is
\begin{eqnarray}
&&\frac{1}{2\tau_{\rm c}} \frac{\partial Q(r)}{\partial t} =
\left[V(0) - V(r) + \frac{1}{\pi \rho_{\rm i} \mu_{\rm in} \tau_{\rm c}} \left( Q(0) + \frac{B_0^2}{6}\right) \right] \times \nonumber \\
&&\left(Q'' + \frac{4Q'}{r}\right) - V' Q' - \frac{1}{r}\left(4V' + r V''\right)
\left(Q+\frac{B_0^2}{6}\right)
\end{eqnarray}
where $V(r)$ is a paired correlation function of turbulent
velocities  of  neutral gas
\begin{eqnarray}
&&\langle v_i(\textbf{x},t) v_j(\textbf{x+r},\bar{t}) \rangle =\nonumber\\
 &&\left( 2 V(r) \delta_{ij} + r V'(r) (\delta_{ij} - \frac{r_i
r_j}{r^2} ) \right) \tau_{\rm c} \delta(t-\bar{t})
\end{eqnarray}
Here $f'$ means $df/dr$ etc.


\begin{figure}
\centering
\includegraphics[width=0.6\textwidth]{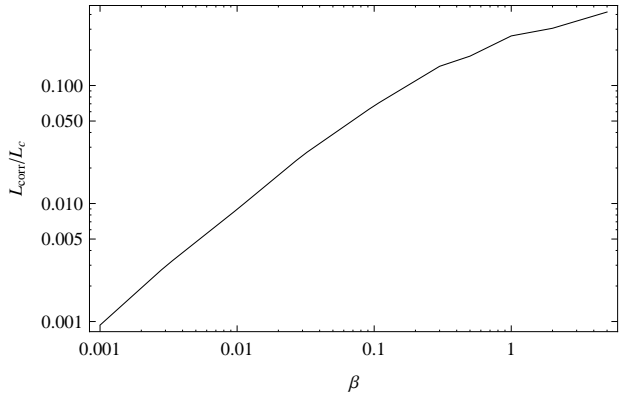}
\caption{Correlation length of magnetic fluctuations $L_{corr} /
L_c$ against the value of parameter $\beta$ }\label{kiselev}
\end{figure}

Numerical calculations of  \citet{kiselev13}  showed that for
small values of $B_0$ such as  the parameter $\beta = B_0 /
(\tau_c \rho_i \mu_{in} v^2)^{0.5}<<1 $, the correlation length of
magnetic fluctuations, $L_{corr}$, is much smaller than the cloud size $L_c$
(see Fig. \ref{kiselev}) i.e. the energy of magnetic fields is
concentrated at scales much smaller than the cloud size, and the
structure of magnetic fields  lines is strongly tangled. For
typical clouds  $L_c = 1 pc$, $v = 1$ km s$^{-1}$, $n_H =
10^{4}$cm$^{-3}$, $n_i = 10^{-3}$cm$^{-3}$ - neutral and ion
densities.  For the gas density in the range from $10^2$ to $10^6$ cmC$^{-3}$ the magnetid field strength is in the limits from $B_0 \sim 30 \, \mu G$ to almost $1$ mG \citep[see][] {Crutcher10}. Then
we obtain $\beta \leq 0.1$ and for  the diffusion coefficient of
CRs in the clouds we have $D_c\sim 10^{26}$cm$^2$s$^{-1}$ i.e.
$D_c<<D_0$, where $D_0\simeq 10^{28}-10^{29}$ cm$^2$s$^{-1}$ is the spatial diffusion coefficient in the intercloud medium \citep[see e.g.][]{ber90}.

\section{Hydrogen and iron ionization by CR protons inside molecular clouds}
Using these parameters of CR propagation in molecular clouds we
estimate the rate of ionization by protons for the Arches
molecular complex as an example. The question is whether the
observed iron emission from the Arches cluster region is produced:
\begin {enumerate}
\item by background CRs which ionise  the diffused molecular gas in
the GC,
\item or by CRs from local sources nearby the cloud.
\end{enumerate}

For these two cases CR distribution, ${N}_p(E,x)$, inside the
cloud can be described by the equation
\begin{equation}\label{pr_cl}
 \frac{\partial}{\partial E}\left( b(E) N_p\right) -  D_c\frac{\partial^2}{\partial x^2} N_p
 = 0\,,
\end{equation}
whose  rate of energy loses is determined by the ionization ,
which for subrelativistic protons can be taken in the form
\begin{equation}
\left(\frac{dE}{dt}\right)_i\equiv b(E)=-\frac{2\pi n
e^4}{mc\beta(E)}\ln\Lambda(E)\,.
\end{equation}
where $m$ is the electron mass, $v$ is the proton velocity,
$\beta=v/c$,  $\ln\Lambda$ is the Coulomb logarithm, and $x$ is
the coordinate from the cloud surface  to the cloud center.

The boundary conditions on the cloud surface ($x=0$) for these
case are different:
\begin{equation}\label{bn}
N|_{x=0}=n_c(E)
\end{equation}
in the  case 1. Here  $n_c(E)$ is the density of background
protons, and
\begin{equation} \label{bq}
\left.\frac{\partial N}{\partial x}\right|_{x=0}=Q(E)
\end{equation}
in the  case 2, where $Q(E)$ is the luminosity of local CR sources
at the cloud surface. In the both cases the second boundary
condition is taken as
\begin{equation}
N|_{x=\infty}=0
\end{equation}

For the time of energy losses
\begin{equation}
\tau_c(E,E_0)=\int\limits_{E_0}^E\frac{dt}{b(t)}\,. \label{t_loss}
\end{equation}
where $E_0$ is the initial energy of a particle, Eq. (\ref{pr_cl})
can be transformed to the standard diffusion equation
\begin{equation}
 \frac{\partial}{\partial \tau}\left(  N^\prime\right) -  D_c\frac{\partial^2}{\partial x^2}
 N^\prime
 = 0\,,
 \end{equation}
where $N^\prime=\mid b(E)\mid N_p$. Solutions for these two cases
can be obtained with the Green function presented e.g. in
\citet{mors}.

In the simplest case of free penetration the ionization rate in
the Arches cloud can be estimated from the observed 6.4 keV flux
which is $F_{6.4}=8.5\times 10^{-6}$ ph cm$^{-2}$s$^{-1}$. As in
\citet{tati12} we take for the estimates: the total mass of the
gas equaled $6\times 10^4$ M$_\odot$, and the gas column density
 about $L_{H_2}\sim 10^{23}$cm$^{-2}$.
With these parameters we obtain that the ionization rate inside
the cloud is about $\zeta_c \leq  10^{-13}$ s$^{-1}$, i.e. the
density of CR protons inside the Arches cloud should be in about
30 - 100 times higher than the CR background in the GC derived
from the $H_3^+$ absorption lines. Thus, the 6.4 keV flux from the
cloud is provided by CRs from nearby local sources.

If fluctuations of magnetic field prevent CR free penetration into
the cloud, the particles fill the region about
$x_c\sim\sqrt{D_c\tau(E)}$ nearby the cloud surface. Here $D_c$ is
the diffusion coefficient inside the cloud. As numerical
calculations show, the estimate of ionization rate inside the
Arches cloud  is absolutely the same as for the case of uniform CR
density there, if $D_c=10^{28}$ cm$^2$s$^{-1}$, i.e. protons fill
almost uniformly the cloud volume. But if $D_c$ is as small as
$10^{25}$ cm$^2$s$^{-1}$, the  CR density is strongly nonuniform
in the cloud. The local ionization rate near the cloud surface is
about $\zeta\sim 10^{-12}$s$^{-1}$ and drops to zero away from the
surface. The ionization rate averaged over the cloud volume (that
is observed from the IR absorption lines) is in this case about
$\bar{\zeta}\sim 5\times 10^{-14}$s$^{-1}$.

Thus we conclude that local sources of CR protons/nuclei generate
the observed 6.4 keV flux from the Arches molecular complex. The
required luminosity of local CR sources (the lower limit in the
thick target approximation) can be estimated as
\begin{equation}
L_{CR}(E>10~\mbox{MeV})\geq  4\pi R^2\frac{F_{6.4}}
{\sigma_{Fe}\eta v}\left(\frac{dE}{dt}\right)_i\sim
10^{39}\mbox{erg s$^{-1}$}\,, \label{cr_lum}
\end{equation}
where $F_{6.4}$ is the flux of 6.4 keV line from Arches, $E_X=6.4$
keV, $R=8$ kpc is the distance between the GC and Earth, $\eta$ is
the iron abundance, $v$ is the proton velocity,
 $\sigma_{Fe}$ is cross-sections of  iron ionization,
 respectively. The spectrum of CRs was supposed to be power-law,
 $E^{-2}$. We notice that this estimate is correct for the
 boundary condition (\ref{bq}) but not for (\ref{bn})

 The estimate (\ref{cr_lum}) is in agreement with obtained by
 \citet{tati12}. We notice that a quite high luminosity of
 subrelativistic CRs is required in order to produce the line flux
 from the region of $\sim 3$ pc diameter.

\section{Ionization by CR electrons in the intercloud medium, stationary component}

We investigate separately the electron origin of the stationary component of the 6.4 keV line
from the diffuse molecular gas in the GC region
and the origin of the time-variable component from dense molecular
clouds.

As \citet{yus07}
showed, there was a correlation between the spatial distribution of
 radio emission and the molecular gas in the GC. Parameters of the electron spectrum in the GC region can be
estimated from the observed flux of radio emission produced by
synchrotron losses of relativistic electrons.  \citet{yus3}
assumed that these electrons ionize diffuse hydrogen gas and generate also the 6.4 keV line.
As follows from \citet{oka05} the rate of ionization is
almost constant throughout the GC region; this means an almost
uniform distribution of the electrons there.

\citet{yus3} derived parameters of these electrons from the observed  nonthermal radio flux from the GC region $2^\circ\times
0.85^\circ$. The flux is about  $\Phi_\nu\simeq 2450$ Jy  at 325 MHz and the
radio spectrum is $\nu^{-0.25}$ at frequencies $\nu<3.3$ GHz. They
assumed that the spectrum of radioemitting electrons
could be extrapolated into the region 100 keV - 1 GeV, and just
the electrons with $E<1$ GeV ionized the GC molecular gas with the
rate $\zeta\sim (1-3)\times 10^{-15}$ s$^{-1}$;  this about the value derived by
\citep[see][]{oka05}. Then the question is whether these electrons
can provide the diffuse 6.4 keV emission in the GC as observed by
\citet{uchiyama2012}.

Parameters of the electron spectrum
in the form $N_e(E)=KE^{-\gamma}$ can be estimated from the
observed radio flux
$\Phi_\nu$ \citep[see e.g.][]{ber90}
 \begin{equation}
  \Phi_\nu=a(\gamma){e^3\over{mc^2}}\left({{3e}\over{4\pi m^3c^5}}\right)^{\frac{\gamma-1}{2}}
  \frac{VKH_\perp^{\frac{\gamma+1}{2}}}{R^2}\nu^{-\frac{\gamma-1}{2}}
  \label{iap}
  \end{equation}
  where the constant $a(\gamma)=0.15$ for the electron spectral index $\gamma=1.5$, $V$
is the volume of $2^\circ\times 0.85^\circ$ GC region, $R\simeq 8$
kpc is the distance from the Sun to the GC, $H$ is the magnetic
field strength in the GC. The values of $\nu$ and $\Phi_\nu$ are
given above. For the magnetic field strength $H=10^{-5}$ G
electrons with the energy $E\simeq 2.7$ GeV radiate at the
frequency $\nu =325$ MHz.

However, the procedure of extrapolation is not trivial because the
electron spectrum is determined by processes of their injection
and energy losses. For the GC region filled with the diffuse
molecular gas ($n=100$ cm$^{-3}$) and the magnetic field
($H=10^{-5}$ G) the synchrocompton losses are significant for
energies above 250 GeV. Below this energy electrons lose their
energy by bremsstrahlung and  then by ionization at $E$ below 350
MeV. Therefore, for $E<250$ GeV we take the rate of energy losses
in the form
\begin{equation}
 \frac{dE}{dt}\equiv
b_0(E)=\left(\frac{dE}{dt}\right)_{i}+\left(\frac{dE}{dt}\right)_{br}\label{loss}
\end{equation}

\begin{figure}
\centering
\includegraphics[width=0.6\textwidth]{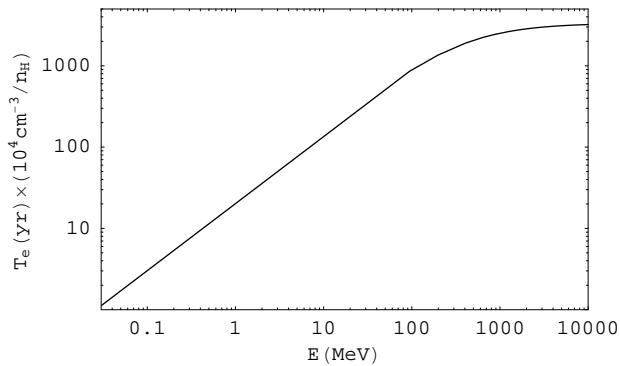}
\caption{The lifetime of electrons  for the gas density
$n_H=10^4$cm$^{-3}$. The lifetime of electrons is determined by
ionization losses up to $E=350$ MeV. Above these energies
electrons lose their energy by bremsstrahlung} \label{lifetime}
\end{figure}

The  lifetime of electrons $T$ in the GC region is shown in
Fig. \ref{lifetime}.

\begin{figure}
\centering
\includegraphics[width=0.7\textwidth]{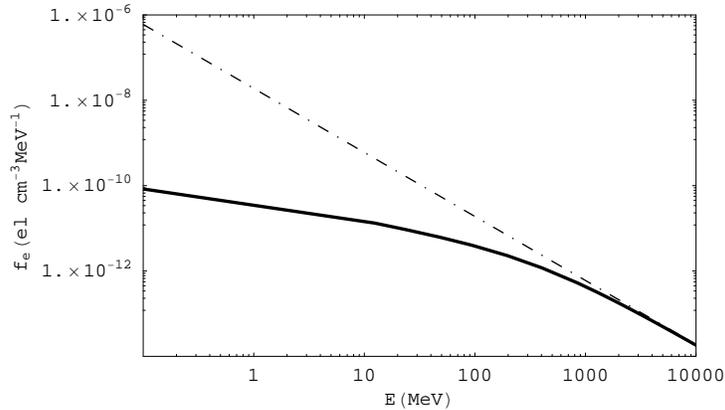}
\caption{The spectrum of electrons in the GC (solid line) as
derived from Eq. (\ref{equ}). By the dashed-dotted line the
extrapolation of electron spectrum into the range below 1 GeV
with the spectral index $\gamma=1.5$ derived from the radiodata
by \citet{yus3}.} \label{spectrum}
\end{figure}

 In the most favorite case for the electron model of the GC ionization, electrons
 lose all their energy in the GC i.e. processes of electron
 escape from there are  insignificant. Then the kinetic equation for the electron
distribution function $f_e$ in the GC reads as
\begin{equation}
\frac{d}{dE}\left(b_0(E)f_e\right)=Q(E) \label{equ}
\end{equation}
If  the source function $Q(E)$ is power-law, $Q(E)=Q_0E^{-1.5}$ as
derived from the radio data (the bremsstrahlung losses do not
change the injection spectral index), then from Eqs. (\ref{iap}) -
(\ref{equ}) we obtain the electron spectrum in the GC as
shown in Fig. \ref{spectrum} by the solid line. The extrapolation
of \citet{yus3} is shown in Fig. \ref{spectrum} by the dashed-dotted line.
One can see that both the spectra are the same in the range of
radioemitting electrons, i.e. above 1 GeV but below this energy
they differ strongly from each other.  Therefore, it is not
surprising that the ionization rate calculated from \citet{yus3}
approximation is about $\zeta\simeq  10^{-15}$ s$^{-1}$, while the
ionization rate calculated from the spectrum derived from the
kinetic equation (\ref{equ}) is only $\zeta\simeq 10^{-17}$
s$^{-1}$.

 In
this respect it is difficult to imagine that ionization by
electrons can provide the necessary ionization rate of hydrogen and
the diffuse flux of the 6.4 keV line in the interstellar GC region
because the intensity of electrons in the energy range $<1$ GeV is strongly
depleted by the ionization losses. Therefore we conclude that the
electrons are unable to provide the necessary rate of ionization
in the GC region, and especially the necessary intensity of the
6.4 keV line from there \citep[see][]{dog13}.

\section{Ionization by CR electrons in molecular clouds, time-variable emission}

Low energy electrons lose their energy effectively by ionization
losses. Thus, the lifetime of 100 keV electrons is about 10 years
(see Fig. \ref{lifetime}). This is about of the observed time
variability of the 6.4 keV line and continuous emissions from some
of the GC molecular clouds. In this respect the electron  model
can be considered as quite reasonable for local production of a
flux of the 6.4 keV line from some of molecular clouds
\citep[see][]{yus4} and even for interpretation of time
variability of this flux as assumed by \citet{yus3}. The question
is whether this model is able to explain  the observed
simultaneous variability of the 6.4 keV emission with the
characteristic time about $\geq 10$ yr for the clouds which are at
distances about 100 pc from each other as follows from the
analyses of  \citet{ponti10,ryu12} and \citet{clavel13}. It seems
incredible that this simultaneous variability is due to
independent local sources of electrons nearby the clouds. From the
variability correlation we  assume that there is a single
time-variable source of electrons in the GC. As a source of
electrons we can assume a SNR which emits electrons for about 3000
yr.

First, we mention  an advantage of the photon model (see Introduction).
Photons propagate along  straight lines with the light speed. If a source
 of photons emits them for a finite period, then these photons provide
 an increase and decrease of the 6.4 keV line flux from a molecular cloud when
 the leading and back fronts of the photon flux cross the cloud. Since the
 effect of photon scattering is negligible the photons can provide the same short-time
 simultaneous variations of the line emission in distant from each other regions long after their ejection.

  Unlike photons,  CR propagation in the interstellar
magnetic fields is described as diffusion due to scattering on magnetic fluctuations \citep[see e.g.][]{ber90}.
Even for the diffusion
coefficient as large as $D=10^{29}$ cm$^2$s$^{-1}$ it takes about
$\geq 10^4$ yr in order to reach a distance about 100 pc from the
source. In the gas with the density $n=100$ cm$^{-3}$ only
electrons with the initial energy $E\geq 10$ MeV can propagate
over this distance (see Fig. \ref{lifetime}).

We can present a source of electrons in the form
\begin{equation}
Q(E,\textbf{r},t)=KE^{-\gamma}\delta(\textbf{r})\theta(t)\theta(T-t)\,,
\end{equation}
where $K$ is a constant, for the period of source activity we take
$T=10$ yr that is most favourable injection time for the electron
model,
 and $\theta(x)$ is the  Heaviside
step-function.

The non-stationary diffusion equation for electron has the form
\begin{equation}
\frac{\partial f_e}{\partial t}-\nabla D\nabla f_e+\frac{\partial
}{\partial E}\left(b_0(E)f_e\right)=Q(E,\textbf{r},t)
\label{nst_equ}
\end{equation}
where $b_0(E)$ is given by Eq. (\ref{loss}). The general solution
of this equation was obtained by \citet{syr59} \citep[see
also][]{ber90}. This solution reads as
\begin{eqnarray}
&&f_e(E,\textbf{r},t)=\frac{K}{\mid
b_0(E)\mid}\int\limits_{E}^{E_{max}}dE_0
\frac{E_0^{-\gamma}}{(4\pi
D\tau(E,E_0))^{3/2}}\times\nonumber\\
&&\exp\left[-\frac{\textbf{r}^2}{4
D\tau(E,E_0)}\right]\theta(\triangle
t+T-\tau(E,E_0))\theta(\tau(E,E_0)-\triangle t) \label{ep_sp}
\end{eqnarray}
where $\triangle t$ is the time after the moment when the source
ceases particle ejection, $b_0(E)$ is described by Eq.
(\ref{loss}),  $E_{max}$ is maximum energy of ejected electrons
and,
\begin{equation}
\tau(E,E_0)=\int\limits_{E_0}^E\frac{dz}{b_0(z)}\,.
\end{equation}

From the solution (\ref{ep_sp}) we can calculate a time-variable flux of the 6.4 keV
line from a cloud which is at the distance 100 pc from the electron source
\begin{equation}
F_{6.4}^e(r=100~\mbox{pc},t) \simeq \eta_{Fe}n_HV\int\limits_{E_e}dE_e \sigma_{e}^{\rm Fe}(E_e)  f_e(E_e,r=100~\mbox{pc},t)v_e
\end{equation}
where $\sigma_{e}^{\rm Fe}$ is the cross-section of iron
ionization  by electrons \citep[see][]{tati03}, $\eta_{Fe}$ is the
iron abundance,  and $V$ is the cloud volume.

\begin{figure}
\centering
\includegraphics[width=0.6\textwidth]{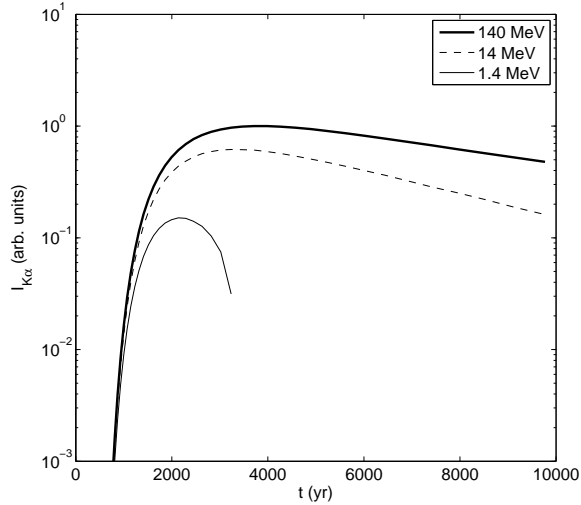}
\caption{The flux of the 6.4 keV line at the distance 100 pc from
the source for different maximum energy of ejected
electrons $E_{max}$.} \label{evol}
\end{figure}

The results of calculations are shown in  Fig. \ref{evol} for
different values of $E_{max}$. As one can see from  the figure the
characteristic time  of the line variations is  $\geq 1000$ yr
and does not reproduce the situation presented e.g. in
\citet{ponti10}.
In this respect we find that  the  model of ionization by
electrons has problems to explain the observed time variations of
the 6.4 keV flux from the GC clouds.

\section{Conclusion}
The conclusions can be itemized as follows:
\begin{itemize}
 \item We conclude that CRs may provide a background level of the
6.4 keV emission in the GC molecular clouds even when the front of
photons ejected by Sgr A* leaves them. The density of CRs in the
GC region can be estimated from the ionization rate of hydrogen in
the GC which is derived from the observed IR absorption lines of
$H_3^+$. The expected background level of the line emission from e.g. the cloud Sgr B2 depends on the spectral index of subrelativistic CRs and it is
quite small for steep CR spectra, $<1$\% of its maximum in 2000.
If in several years this background level is higher than this
value, then the reason may be due to ejection of protons by accretion
onto the central black hole or other local sources of CRs or
photons;
\item Local sources of CRs may provide an excess of the 6.4 keV line emission in some
of the GC molecular clouds. The required density of CRs  depends
strongly on processes of particle penetration into the clouds. The
analysis of magnetic fields  inside the clouds shows that strong
small scale magnetic fluctuations may excite there by   the
turbulence of neutral gas, which prevents free penetration of
charged particles into the clouds. Therefore CRs distribution
there is strongly nonuniform. However, the CR luminosity of local
sources needed for e.g. the observed 6.4 keV flux from the  Arches
complex is independent of how CRs are distributed there, and
equals about $10^{39}$ erg s$^{-1}$;
\item Our earlier analysis  \citep{dog13} showed that ionization of the diffuse hydrogen is provided
by subrelativistic protons  while the diffuse 6.4 keV line in the
GC is generated by X-ray photons emitted by Sgr A*. An alternative
interpretation was suggested by \citet{yus3} who assumed that the
ionization is provided by background electrons with $E<1$ GeV
whose spectrum was extrapolated from the radio data. We showed
that the model of ionization by electrons is problematic because
the intensity of electrons in the range $<1$ GeV is strongly depleted
by  ionization losses. However, we cannot exclude (rather
arbitrarily) an upturn of the injection spectrum of electrons  in
the range below 1 GeV . Then one may expect a higher ionization
rate in the GC than estimated in section 5. In any case, the local
electron spectrum near a  very compact region nearby the cloud
GO.13-013 derived by \citet{yus4} from the radio data from 74 to
327 MHz, $\sim E_e^{-3.6}$, is much steeper  than derived by
\citet{yus3} from the diffuse radio emission in the GC. However.
the analysis of \citet{yus4} showed strong spatial variations of
the electron spectral index  in this region, and it is unlikely to
assume this spectrum in the whole GC region;
 \item In our opinion another problem of the electron interpretation is a
 relatively high luminosity of electrons needed for hydrogen ionization
 in the GC region.
 As \citet{yus3} showed the energy
 supply of  electrons necessary for the 6.4 keV line flux from the GC
 is  about $10^{41}$
 erg s$^{-1}$ that is one order of magnitude higher than the total
 CR luminosity in the Galaxy and three orders of magnitude higher
 than the total electron luminosity of the Galaxy \citep[see][]{strong10}.
 It seems that there are no known sources which are able to sustain such an electron
population throughout the Central Molecular Zone (CMZ);
\item We do not exclude that
 local sources of electrons may provide an  excesses of the 6.4 keV line
 emission in some molecular clouds  and even reproduce a relatively short time
 variations of the iron line emission ($\sim 10$ yr) \citep[see][]{yus3,yus4}. However,
 if interpretation of the 6.4 keV time variability of e.g \citet{ponti10, ryu12,clavel13} is correct, then the electron model is
  unable to reproduce   the simultaneous short time variability
of the iron line emission from clouds which are distant from each
other by ~hundred pc.  Alternatively we can speculate that a random
distribution of local electron sources could also provide the
necessary effect of the line variations in different clouds that
are seen by chance.   However, this interpretation seems to be very unlikely.
\item In our opinion  the photon model of \citet{ponti10,
ryu12,clavel13} reproduces naturally these
 spatial and temporal characteristics of the 6.4 keV emission in
 the GC. The photon model has two important advantages: the front
 of primary photons occupies a very extended region  100 yr
 after the explosion,  and its spatial distribution may have  very sharp leading and back fronts  that
 reproduces easily the 10 yr temporal variations of the 6.4 keV
 emission  from GC molecular clouds.

\end{itemize}

\section*{Acknowledgements}
The authors
 are grateful to Farhad Yusef-Zadeh for critical reading of the manuscript and helpful discussions,  and to the referee Roland Crocker, for very useful comments and corrections which helped us much to improve the text.
 VAD and DOC acknowledge support from the
International Space Science Institute to the International Team
216 and the RFFI grants 12-02-00005 and 14-02-00718. DOC
and AMK are supported in parts by the RFFI grant 12-02-31648 and the LPI Educational-Scientific Complex. DOC is  supported by
the Dynasty Foundation. AMK is supported by the RFFI grant 11-02-01021. KSC is supported by a GRF grant from the Hong Kong Government under HKU 701013.

\end{document}